\documentclass[reprint,aps,amsmath,amssymb]{revtex4-2} 
\usepackage[rmargin=2.5cm,lmargin=2.5cm,tmargin=1.5cm]{geometry} 
\usepackage[utf8]{inputenc}
\usepackage{amsmath}
\usepackage{amsfonts}
\usepackage{amssymb}
\usepackage{graphicx}
\usepackage{listings}
\usepackage{color} 
\definecolor{mygreen}{RGB}{28,172,0} 
\definecolor{mylilas}{RGB}{170,55,241}
\usepackage{url} 
\usepackage[normalem]{ulem} 
\useunder{\uline}{\ul}{} 

\begin{document}
\title{Orders-of-magnitude reduction in photonic mode volume by nano-sculpting} 

\author{Rasmus E. Christiansen$^{1,3}$}
\email{Corresponding email: raelch@dtu.dk}

\author{Jesper Mørk$^{2,3}$}

\author{Ole Sigmund$^{1,3}$}

\affiliation{$^{1}$Department of Civil and Mechanical Engineering, Technical University of Denmark, Nils Koppels Allé 404, DK-2800 Kgs. Lyngby, Denmark} 
\affiliation{$^{2}$Department of Electrical and Photonics Engineering, Technical University of Denmark, Ørsteds Plads 343, DK-2800 Kgs. Lyngby, Denmark} 
\affiliation{$^{3}$NanoPhoton - Center for Nanophotonics, Technical University of Denmark, Ørsteds Plads 345A, DK-2800 Kgs. Lyngby, Denmark.}

\begin{abstract}
\noindent Achieving strong light-matter interaction is important for studying and exploiting several physics phenomena. The light-matter interaction strength depends on the optical field intensity in the interaction region, often measured by the Purcell factor, which for a single emitter is proportional to the spectral confinement, quantified by the cavity quality factor $Q$, and inversely proportional to the spatial localization of light, quantified by the optical model volume $V$, $F \propto \frac{Q}{V}$. While plasmonic (metallic) devices can support extreme spatial light confinement, ohmic losses reduce the cavity lifetime, thereby limiting the achievable spectral confinement. It is therefore of both practical and fundamental interest to explore the potential for achieving extreme spatial light confinement in (near) loss-less dielectric environments. 

Employing topology optimization we explore the limits of spatial light confinement in dielectric environments when allowing for three-dimensional sculpted dielectric nanostructures. Here we discover structures supporting optical modes that are concentrated in material (air) with mode volumes that are three (four) orders of magnitude below the so-called diffraction limit, $V_{\textbf{r}_0} \approx 4 \cdot 10^{-4} \left[\lambda/(2 n)\right]^3 \left( V_{\textbf{r}_0} \approx 3 \cdot 10^{-5} \left[\lambda/2\right]^3\right)$. Remarkably, we further discover that encapsulating the nanostructure by ellipsoidal shells enables seemingly unbounded enhancement of the mode quality factor ($Q > 10^8$ demonstrated numerically) leading to theoretical Purcell factor enhancement above $10^{11}$. It is established how $V_{\textbf{r}_0}$ and $Q$ depend on the choice of material platform, device volume, minimum feature size and the number of shells. \textcolor{black}{Finally a study of sensitivity towards geometric variations is presented, revealing robust behaviour under a range of perturbations.}

\end{abstract}

\maketitle


\section{Introduction}

The decay rate of an emitter (e.g., an atom) is enhanced by the so-called Purcell factor if the emitter is placed in a cavity that spatially concentrates light and stores it for an extended period of time. The Purcell factor is proportional to the ratio between the spectral and spatial confinement of the light at the emitter position, $r_0$, i.e. $F \propto \frac{Q}{V_{\textbf{r}_0}}$ \cite{PURCELL_1946}. Here, $Q$ denotes the cavity quality factor and $V_{\textbf{r}_0}$ the single-emitter mode volume associated with the electromagnetic resonance of the cavity in which the emitter is embedded. Several potentially ground-breaking applications depend on achieving strong light-matter interaction \cite{Notomi_2010}, e.g. realizing efficient single-photon sources for quantum technology \cite{DING_ET_AL_2016} or nano-scale lasers and light-emitting diodes \cite{YAMAMOTO_ET_AL_1991,LAU_ET_AL_2009,Surh_et_al_2010,MORK_YVIND_2020}. While some applications are able to rely on large $Q$-values to achieve sufficient interaction strength, other applications (like optical switches) require high operating speeds and thus large optical bandwidths; are footprint-constrained; or are otherwise spectrally limited, all requirements that bound $Q$ from above. Such applications must instead exploit strong spatial confinement of the field, i.e. small mode volumes, to increase $F$. Extreme spatial confinement may be achieved using plasmonic (metallic) structures, at the cost of significant energy loss to heating and limited spectral confinement ($Q<100$)\cite{KURGIN_2015}, both being unacceptable conditions for many applications. \textcolor{black}{A potential avenue for significantly reducing the ohmic energy loss may be to exploit hybrid metal-dielectric structures. This idea has been theoretically shown to maintain a high-degree of spatial field confinement while significantly improving radiative efficiency due to lower losses in cylindrical bowtie-like nano-antenna\cite{HOOTEN_ET_AL_2021}. }

\textcolor{black}{Owing to the importance of achieving strong light-matter interaction, numerous works have been dedicated to both the design and experimental realization of devices supporting ultra-high quality factors and small mode volumes. A few examples representing the state-of-the-art are; the experimental achievement of a $Q$-factor exceeding 11 million in a silicon photonic crystal cavity, with a mode volume of $V_{\textbf{r}_0} \approx 8\left[\frac{\lambda}{2n}\right]^3 $ \cite{ASANO_ET_AL_2017}; the achievement of a mode volume of $V_{\textbf{r}_0} \approx 2.64 \cdot 10^{-3} \left[\frac{\lambda}{2}\right]^3$ in a nanoscale disk resonator sustaining metal-insulator-metal plasmons with a $Q$-factor of 16 \cite{KUTTGE_ET_AL_2010}. A hybrid metal-semiconductor approach achieving small mode volumes of $V_{\textbf{r}_0} \approx 1 \cdot 10^{-3} \left[\frac{\lambda}{2}\right]^3$ along with moderately high $Q$-factor of 800 \cite{CONTEDUCA_ET_AL_2017}. Utilizing a whispering gallery mode resonator, a $Q$-factor of $6 \cdot 10^{10}$ has been obtained within a millimeter-scale cavity \cite{GRUDININ_ET_AL_2006}.}

Achieving $V_{\textbf{r}_0} < 1 \left[\frac{\lambda}{2 n}\right]^3$ with light confined inside the solid using only dielectric materials was long deemed unattainable, but recent works have demonstrated this to be possible through so-called Extreme Dielectric Confinement (EDC) both numerically and experimentally  \cite{Robinson_Lipson_2005,Gondarenko_Lipson_2006,LIANG_JOHNSON_2013,CHOI_ET_AL_2017,Hu_Weiss_2018,WANG_2018,Mignuzzi_Sapienza_2019,Isiklar_2022} in platforms made of silicon $V_{r_0} \approx 0.08 \left[\lambda/(2 n_{\text{Si}})\right]^3$ \cite{Albrechtsen_2022}, and indium phosphide, $V_{r_0} \approx 0.26 \left[\lambda/(2 n_{\text{InP}})\right]^3$ \cite{Xiong_2024}. These achievements have opened a new avenue for enhancing light-matter interaction in near-lossless dielectric environments capable of supporting high $Q$-values. 

\begin{figure*}[!]
	\centering
	{
		\includegraphics[width=0.98\textwidth]{Fig1.png} \caption{\textbf{(a)} Planar 2D-patterned dielectric cavity (gray). \textbf{(b)} 3D-sculpted dielectric cavity. \textbf{(c)} Axisymetric 3D-sculpted dielectric cavity. Each panel include a zoom of a quarter of the geometry centered at $\textbf{r}_0$ showing the eletric field magnitude (inferno colormap) of the air-confined mode and lists the associated single-emitter mode volume. \label{FIG:2D_MembraneConfinement}}
	}
\end{figure*} 

Until now, exploration of the EDC phenomenon has been restricted to two-dimensionally (2D) patterned planar devices relying on few-nanometer-sized geometric features to facilitate strong in-plane field confinement (Fig.~\ref{FIG:2D_MembraneConfinement}a). As state-of-the-art nanofabrication capable of realizing features at the few-nanometer scale is limited to patterning 2D-layers of material \cite{GrigorescuHagen_2009}, such a restriction is sensible. However, considering technological advances in three-dimensional (3D) nanoscale fabrication, where the 3D-printing of structures with features in the hundred nanometer range is approaching routine \cite{Chen_2021,TrubyLewis_2016}, realizing 3D-sculpted geometries with features at the few-nanometer scale may become feasible in the not too distant future. This may possibly be realized using technologies based on implosion fabrication \cite{Oran_2018}, ice-lithography \cite{QIU_2019} or focused ion-beam milling \cite{LeviSetti_1974} coupled with advanced nano-assembly technologies \cite{Babar_Stobbe_2023}. \textcolor{black}{In order to exploit the extremely small mode-volume for light-matter interaction, a nanoscale emitter must be embedded in the system such that it overlaps with optical hotspot of the cavity \cite{XIONG_ET_AL_2025}. A deterministic approach for placing localized emitters, such as quantum dots in the hotspot of a nanocavity, remains an unsolved challenge. However, one option for embedding emitters is using ion doping of the host material (e.g. erbium doping of silicon or defects in silicon), relying on the chance location of a single emitter in the hotspot of the cavity. Using this approach and an appropriately chosen erbium concentration, a significant Purcell enhancement was recently demonstrated in a photonic crystal cavity \cite{GRITSCH_ET_AL_2023}. Another approach is to couple the emitter placement with the concept of self-assembled nanostructures \cite{Babar_Stobbe_2023} to realize the few nanometer central bridge/gap with an embedded emitter.}

In this paper we demonstrate the theoretical and numerical possibility of realizing orders of magnitude improvement of the $Q$-over-$V$ ratio by allowing three-dimensional sculpting of the electromagnetic environment using inverse design by Topology Optimization (TopOpt) \cite{BENDSOE_KIKUCHI_1988,CHRISTIANSEN_SIGMUND_COMSOL_2020}, which has garnered significant interest for nano-photonic applications in recent years \cite{MOLESKY_2018,JENSEN_SIGMUND_2011}. \textcolor{black}{This clearly demonstrates the benefit of exploiting the third spatial dimension in device design.} First, a study of the full 3D design problem reveals \textcolor{black}{new EDC geometries, shaped to enhance out-of-plane light confinement by supporting a new mode-shape near $\textbf{r}_0$ (compare inserts in Fig.~\ref{FIG:2D_MembraneConfinement}a and Fig.~\ref{FIG:2D_MembraneConfinement}b) leading to} a reduction in the achievable $V_{\textbf{r}_0}$ with increasing 3D design freedom (Fig.~\ref{FIG:MIN_VR0_3D_STUDY}). Then, by exploiting a fundamental rotational symmetry discovered from the freely 3D-sculpted structures (Fig.~\ref{FIG:FULL3D_FIELD_PROFILE_1600nm}) the problem is reduced to an axisymmetric sculpting problem enabling extreme design resolution, in turn allowing the discovery of a structure supporting an air-confined mode with $V_{\textbf{r}_0} \approx 3 \cdot 10^{-5} \left[\lambda/2\right]^3$ (Fig.~\ref{FIG:2D_MembraneConfinement}c) and another supporting a material-confined mode with $V_{\textbf{r}_0} \approx 8 \cdot 10^{-4} \left[\lambda/(2 n_{\mathrm{Si}})\right]^3$ simultaneously allowing for extreme spectral confinement ($Q>10^8$) by embedding the structure in ellipsoidal shells (Fig.~\ref{FIG:3D_AXISSYMMETRIC_ONION}). A systematic study of the effect of the dielectric material choice, device volume and minimum feature-size on $V_{\textbf{r}_0}$ is presented (Fig.~\ref{FIG:PARAMETER_STUDIES_AXISYM}), \textcolor{black}{revealing a remarkable 1/$n^7$ dependence of the achievable mode-volume on the refractive index at lower index values (Fig.~\ref{FIG:PARAMETER_STUDIES_AXISYM})a}. In the process we identify a silicon structure with 1 nm solid bridge-feature at $\textbf{r}_0$ exhibiting $V_{\textbf{r}_0} \approx 4 \cdot 10^{-4} \left[\lambda/(2 n_{\mathrm{Si}})\right]^3 \approx 1.2 \cdot 10^{-6} \left[\lambda\right]^3$. \textcolor{black}{Finally, a study of the effect of geometric perturbations, mimicking fabrication imperfections, is included, revealing remarkable geometric robustness of both $Q$ and $V_{\textbf{r}_0}$.}

\section{Methods}

We perform minimization of the single-emitter mode volume by sculpting the dielectric environment around $\textbf{r}_0$ based on recasting (approximating) the source-less eigenproblem associated with computing the optical quasi-normal modes\cite{KRISTENSEN_ET_AL_2012} as a driven time-harmonic problem. This has also been done in prior works considering two-dimensional planar structures \cite{LIANG_JOHNSON_2013,WANG_2018}. In contrast to prior works we consider an electromagnetic plane-wave as an excitation source rather than a point dipole at $\textbf{r}_0$. As the objective function to be minimized in our TopOpt-based 3D-sculpting process, we approximate the single-emitter mode volume calculation by replacing the quasi-normal mode with the solution of the driven time-harmonic wave equation yielding the expression,

\begin{eqnarray}
	\Phi = \frac{\int_{\Omega_I} \varepsilon_r(\textbf{r})|\textbf{E}(\textbf{r})|^2 d\textbf{r}}{\varepsilon_r(\textbf{r}_0)|\textbf{E}(\textbf{r}_0)|^2} \approx V_{r_0}. \label{EQN:FIGURE_OF_MERIT_3D} 
\end{eqnarray}

Here $\Omega_I \subset \mathbb{R}^3$ denotes the spatial domain being modeled, $\textbf{E}(\textbf{r})$ the time-harmonic electric field resulting from the external excitation and $\varepsilon_r(\textbf{r})$ the spatially dependent relative permittivity. While eq. \eqref{EQN:FIGURE_OF_MERIT_3D} is used in the inverse design process, all single-emitter mode volumes reported for the optimized structures in this work are computed by solving a source-less eigenproblem using the standard quasi-normal mode approach\cite{KRISTENSEN_ET_AL_2012} verifying accurate computation of $V_{\textbf{r}_0}$ and offering direct comparability to literature.

To simulate the driven time-harmonic electric field in $\Omega_I$ we employ a classical electromagnetics model, assuming one first-order scattering boundary condition to truncate the modeling domain and another to introduce a Gaussian-enveloped ($\sigma = 500$ nm), linearly-polarized, plane wave exciting the structure. \textcolor{black}{Remarkably, while the system is excited asymmetrically using a plane wave, the optimized geometry, and associated eigenmode of interest, exhibit clear axisymmetry near $\textbf{r}_0$, suggesting optimality of this newly identified geometry in the region proximate to $\textbf{r}_0$.}

The initial study is carried out employing a fully three-dimensional (3D) model, while the subsequent studies are carried out using an efficient axisymmetric model assuming field invariance in the azimuthal direction $\textbf{E}(r,\phi,z) = \textbf{E}(r,z)$, i.e. only targeting the lowest order mode in the azimuthal expansion. In the axisymmetric model the plane-wave excitation used in the 3D model is replaced by radially propagating wave excitation. The physics model problem is discretized and solved using the finite-element method \cite{BOOK_FEM_JIN} and the inverse design problem is solved using the Method of Moving Asymptotes \cite{SVANBERG_2002}, all implemented utilizing COMSOL Multiphysics \cite{COMSOL61main}. 

The material distribution in $\Omega_I$, constituting the device to be designed, is controlled by a design field $\xi(\textbf{r}) \in [0,1]$, which is iteratively updated to minimize $\Phi$ subject to a length-scale constraint \cite{ZHOU_ET_AL_2014} controlling the smallest possible feature size of the structure. The design field is linked to the physics model by linearly interpolating the relative permittivity $\varepsilon_r(\textbf{r})$ in $\Omega_I$ between a background material ($\xi(\textbf{r}) = 0$) and the device material ($\xi(\textbf{r}) = 1$) after subjecting the design field to a standard smoothing and thresholding procedure \cite{WANG_ET_AL_2011}. All structures are designed starting from a single uniform initial guess ($\xi(\textbf{r}) = 0.5$). Selecting a fixed $\textbf{r}_0$ as our reference point to compute $V_{\textbf{r}_0}$ and enforcing a minimum length scale of the optimized structures ensure that we do not rely on a lightning-rod type effect at the surface to create singularities in the electric field resulting in artificially small mode volumes \cite{Albrechtsen_2022b}. We notice, that while the conventional definition of the mode volume as a position dependent measure of the local vacuum field strength is relevant for point dipoles, spatially extended emitters should employ a more general definition \cite{SALDUTTI_YI_MORK_2024}. For technical details regarding the numerical modeling, optimization problem and choice of hyper-parameters, see Supplementary Information \ref{SEC:SUPPLE_INFO}.

\section{Results and Discussion}

\begin{figure*}[!]
	\centering
	{
		\includegraphics[width=0.98\textwidth]{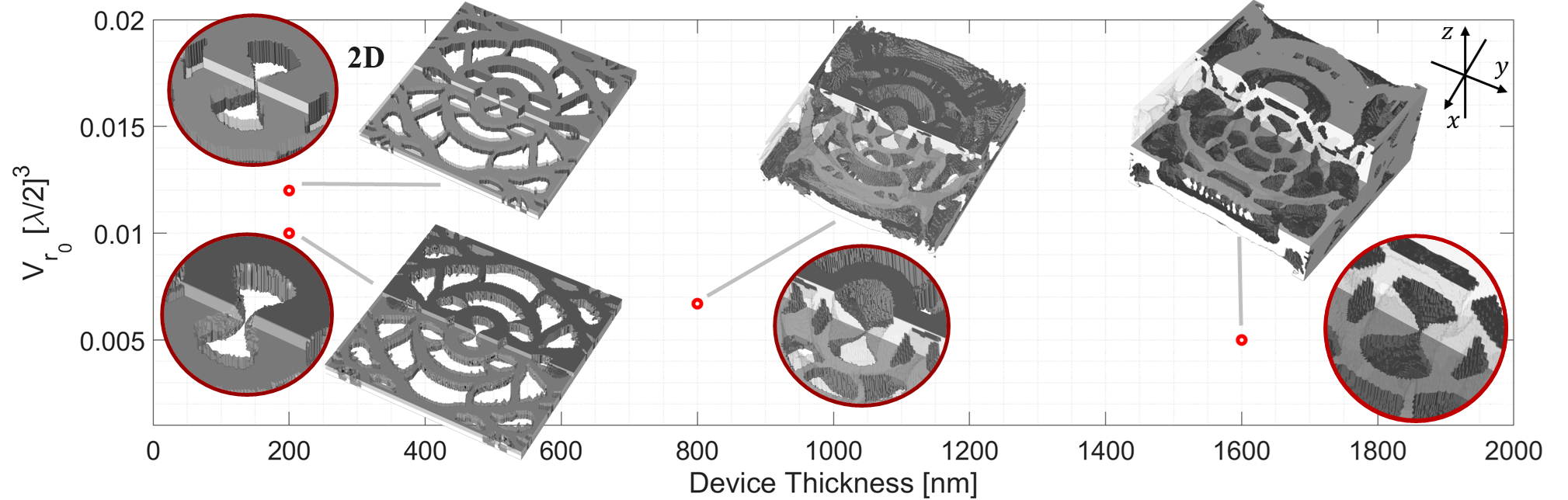} \caption{Single-emitter mode volumes $(V_{\textbf{r}_0})$ for the four optimized structures as a function of allowed device thickness. The minimum feature size allowed for the structure is 10 nm, and the optical intensity is concentrated in the a central air region. The inserts show the optimized geometries for each $(V_{\textbf{r}_0})$-value, with a zoom of the regions surrounding $\textbf{r}_0$. \label{FIG:MIN_VR0_3D_STUDY}}
	}
\end{figure*}

First, we study the effect of introducing 3D-design freedom on the achievable $V_{\textbf{r}_0}$ and on the device geometry. To this end, a planar 2D-patterned membrane (reference) and a set of 3D-sculpted structures are designed for air-confinement at $\textbf{r}_0$. To ensure direct comparability in this investigation, identical (computationally limited) discretizations/resolutions are used for all four designs, with a minimum voxel size of 10 nm. We assume an air background ($n = 1.0$) and a device material corresponding to indium phosphide (InP) at $\lambda = 1550$ nm ($n = 3.17$). The resulting structures and corresponding mode volumes are presented in Fig.~\ref{FIG:MIN_VR0_3D_STUDY}. As seen from Fig.~\ref{FIG:MIN_VR0_3D_STUDY}, the mode volume monotonously decreases with increasing out-of-plane design freedom (device thickness). This reduction of $V_{\textbf{r}_0}$ comes at the cost of a more complex geometry, incorporating a bowtie-like torus-shaped feature that confines the field spatially in both the in-plane and out-of-plane direction at $r_0$. The single-emitter mode volume decreases by a factor of approximately $2$ when increasing the device thickness from $200$ nm to $1600$ nm as the field near the center of the structure is localized further in the out-of-plane direction serving to concentrate it more strongly at $\textbf{r}_0$. This (relatively limited) reduction in $V_{\textbf{r}_0}$ is not fundamental, but a consequence of the limited model resolution possible when treating the full 3D problem numerically, as will become clear in the following studies. By comparing the electric field magnitude $\vert \textbf{E} \vert$ of the confined mode, presented in the zooms in Fig.~\ref{FIG:2D_MembraneConfinement}a and Fig.~\ref{FIG:2D_MembraneConfinement}b, it is directly observable that introducing 3D design freedom leads to tighter out-of-plane field confinement at $\textbf{r}_0$.

\begin{figure}[!]
	\centering
	{
		\includegraphics[width=0.5\textwidth]{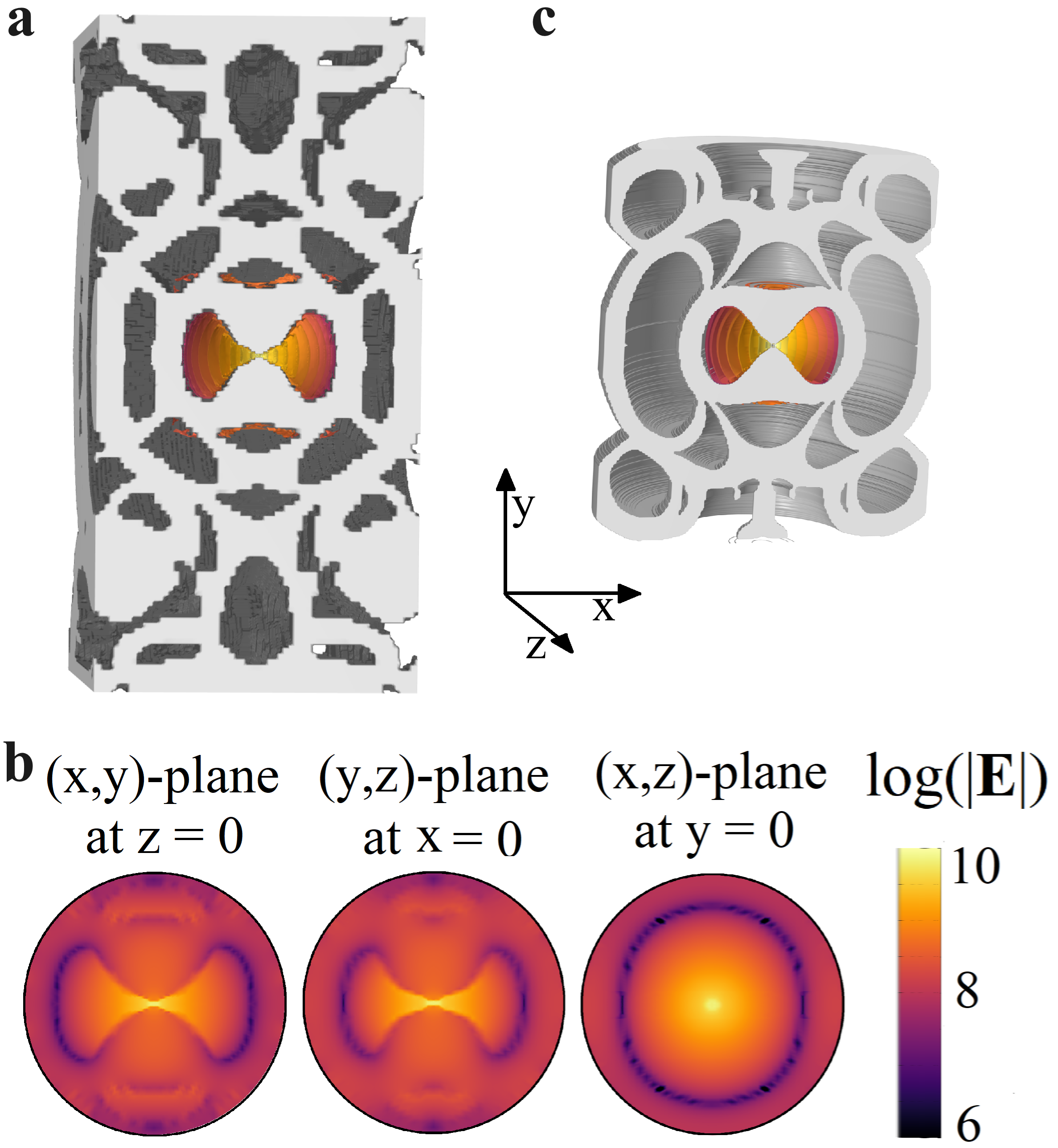} \caption{\textbf{(a)} (y,z)-plane cut at x=0 of 3D-sculpted $1600$ nm thick structure from Fig.~\ref{FIG:MIN_VR0_3D_STUDY} with full 3D design freedom. \textbf{(b)} Electric field magnitude in the (x,y)-, (y,z)- and (x,z)-planes through $\textbf{r}_0$ of targeted mode. \textbf{(c)} (y,z)-plane cut at x=0 through device designed using the axisymmetric model. \textbf{(a)} and \textbf{(c)} include overlays showing $\log_{10}\left(\vert \textbf{E} \vert\right)$ for the targeted mode within a radius of $400$ nm of $\textbf{r}_0$. \label{FIG:FULL3D_FIELD_PROFILE_1600nm}}
	}
\end{figure}

Remarkably, when investigating the 3D-sculpted structures in Fig.~\ref{FIG:MIN_VR0_3D_STUDY}, a bowtie-like torus and other near-axisymmetric features are seen to develop in the region near $\textbf{r}_0$ as the design thickness increases (see the zoomed inserts). Remembering that the structures were designed under asymmetric plane-wave excitation, this suggests a fundamental symmetry in the optimal structure, serving to support a new optical mode capable of supporting stronger light confinement. To investigate this finding further, both the geometry and the field profile of the mode localized at $\textbf{r}_0$ are studied in detail for the $1600$ nm thick structure, see Fig.~\ref{FIG:FULL3D_FIELD_PROFILE_1600nm}a (cross-section of the optimized geometry) and Fig.~\ref{FIG:FULL3D_FIELD_PROFILE_1600nm}b (cross-sections of the electric-field profile). In doing so we find the important result that even though complete 3D-sculpting freedom was allowed in the inverse design process, and even though the design problem is excited by a plane wave propagating through the domain and not by a point dipole at $r_0$, which might inherently promote symmetry due to its symmetric emission pattern, both the optimized geometry and the resulting eigenmode profile exhibit near-axisymmetry (about the y-axis) in the vicinity of $\textbf{r}_0$, suggesting that this configuration is indeed optimal. 

Motivated by this observation we set up an axisymmetric electromagnetic model problem (see e.g. \cite{CHRISTIANSEN_OE_2020b} for an example of electromagnetic TopOpt under axisymmetric assumption) with parameters that are otherwise identical to the full 3D model. We employ this model to sculpt a structure resulting in the geometry shown in Fig.~\ref{FIG:FULL3D_FIELD_PROFILE_1600nm}c. When comparing Fig.~\ref{FIG:FULL3D_FIELD_PROFILE_1600nm}a and Fig.~\ref{FIG:FULL3D_FIELD_PROFILE_1600nm}c similar geometric features and electric field profiles are observed, suggesting that the axisymetric model captures the important design features at a fraction of the computational cost. In this way, the full 3D model is replaced by the axisymmetric physics model, hereby enabling numerical studies at significantly finer geometric resolution and larger overall device sizes. 

\begin{figure*}[!]
	\centering
	{
	\includegraphics[width=0.98\textwidth]{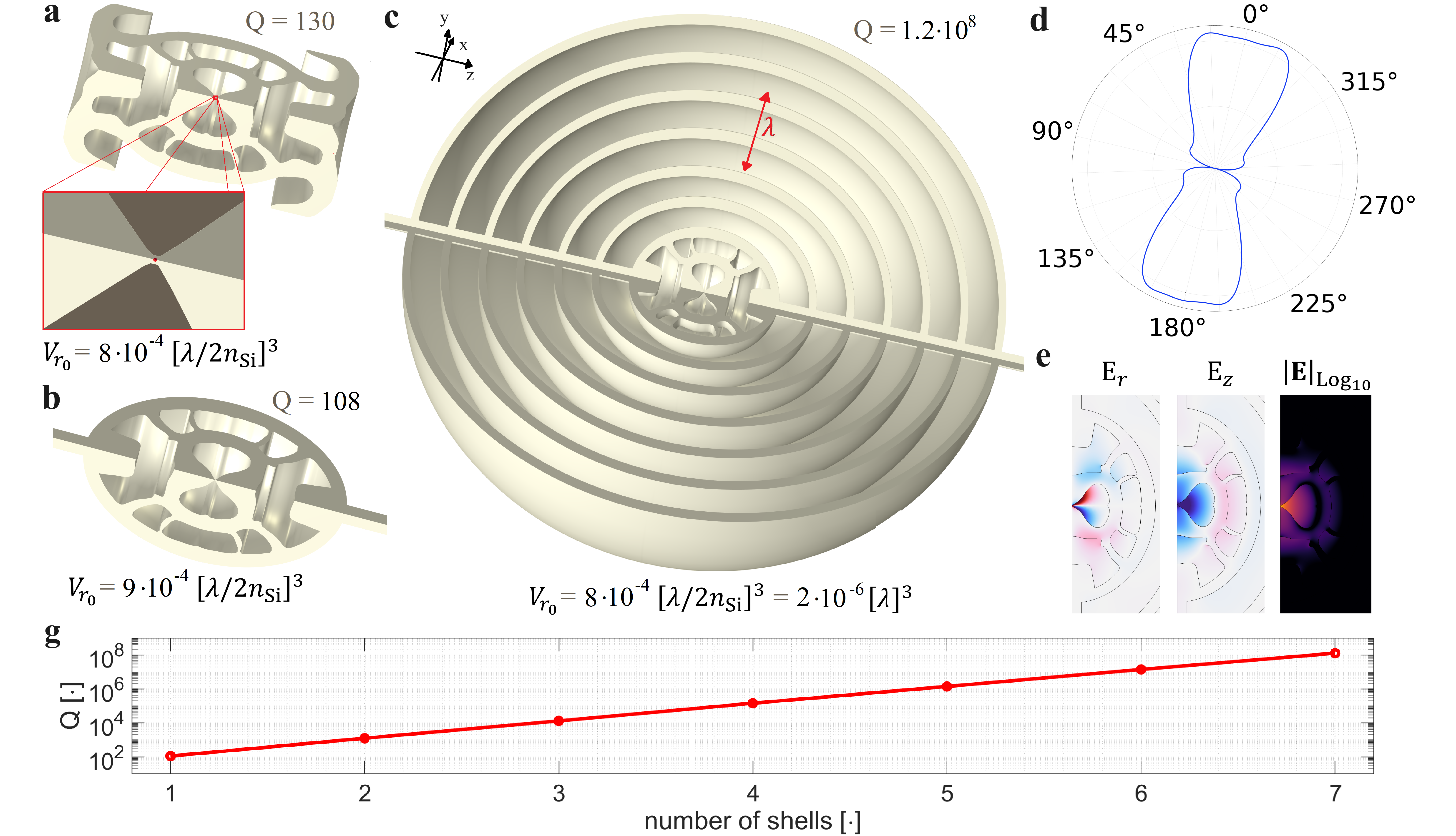} \caption{Quarter revolution of \textbf{(a)} axisymmetric silicon device with a fixed 2 nm diamater central solid bridge-feature. \textbf{(b)} simplified geometry with ellipsoidal shell and connecting rod. \textbf{(c)} Three-quarter revolution of simplified geometry with six additional ellipsoildal shells. \textbf{(d)} Far-field emission pattern in the (x,z)-plane. \textbf{(e)} Electric near-field components and magnitude. \textbf{(g)} Quality factor of the relevant mode as a function of the number of shells. \label{FIG:3D_AXISSYMMETRIC_ONION}}	}
\end{figure*}

Employing the computationally efficient axisymmetric model we first investigate the achievable single-emitter mode volume when confining the field in the air (Fig.~\ref{FIG:2D_MembraneConfinement}c). Here a 1 nm air gap (minimum feature size) is enforced at $\textbf{r}_0$, comparable to what was recently experimentally demonstrated to be fabricable when considering planar 2D-patterned devices realized using self-assembled silicon nanostructures \cite{Babar_Stobbe_2023}. In difference to prior planar structures, this structure is found to support a non-planar but rather axisymmetric mode with $V_{\textbf{r}_0} \approx 3 \cdot 10^{-5} \left[\lambda/2\right]^3$. Thus it is revealed, that through 3D-sculpting it is possible to achieve single-emitter mode volumes nearly five orders of magnitude below the diffraction limit for field confinement in a 1 nm air gap. A more than 50x smaller $V_{\textbf{r}_0}$ than what was reported theoretically for a planar nano-beam \cite{CHOI_ET_AL_2017} with an identical gap size is predicted. 

Next, employing the axisymmetric model we investigate the achievable single-emitter mode volume when confining the field in a solid bridge (cylinder). To this end, we assume a background index of $n = 1.0$ and a device material index of $n = 3.48$ (silicon at $\lambda = 1550$ nm) and enforce a fixed 2 nm diameter central solid bridge-feature. The inversely designed structure obtained under these conditions is presented in Fig.~\ref{FIG:3D_AXISSYMMETRIC_ONION}a. 

This device supports a single-emitter mode volume of $V_{\textbf{r}_0} \approx 8\cdot 10^{-4} \left[\lambda/(2 n_{Si})\right]^3$ for the appropriate eigenmode, two orders of magnitude smaller than the record-low mode volume reported in \cite{Albrechtsen_2022} for a 2D-patterned silicon structure. \textcolor{black}{Carefully studying the profile of the strongly confined mode, and making the non-trivial observation} that the electric near-field profile immediately outside the device geometry is approximately ellipsoidal, a simplification of the geometry is made by replacing the outermost part by an ellipsoidal shell, see Fig.~\ref{FIG:3D_AXISSYMMETRIC_ONION}b. This operation simplifies the geometry at the cost of increasing $V_{\textbf{r}_0}$ by approximately 10\%. In turn we find that the simplification enables systematic control of the mode quality factor through the introduction of additional ellipsoidal shells with parameter-optimized major-axis radii and thicknesses. Simply introducing the shells results in a spatially disconnected, mechanically infeasible, structure. However, remarkably we find that introducing a solid rod along the z-axis, where the emission pattern exhibits a node (Fig.~\ref{FIG:3D_AXISSYMMETRIC_ONION}d), enables us to connect all shells to each other and to the central device, with (almost) no effect on the spatio-spectral confinement. 

Figure~\ref{FIG:3D_AXISSYMMETRIC_ONION}c shows a rendering of the device with six shells and the rod added. This design configuration raises the quality factor from $Q\approx130$ to $Q\approx1.2 \cdot 10^8$ without $V_{\textbf{r}_0}$ increasing. The shells operate as a Bragg-like mirror, each layer increasing $Q$ by approximately one order of magnitude as reported in the graph in Fig.~\ref{FIG:3D_AXISSYMMETRIC_ONION}f. Thus, if the emitter-cavity structure is limited by the linewidth of the cavity rather than homogeneous broadening of the emitter \cite{MORK_LIPPI_2018}, the device shown in Fig. \ref{FIG:3D_AXISSYMMETRIC_ONION}c supports Purcell enhancement of $F \approx 10^{11}$. This is a several orders of magnitude greater enhancement than what was observed for a previously proposed 2D-patterned nanobeam device \cite{CHOI_ET_AL_2017}, an enhancement resulting from both  significantly smaller $V_{\textbf{r}_0}$ and larger $Q$. The far-field profile and the near-field of the device in Fig.~\ref{FIG:3D_AXISSYMMETRIC_ONION}c are visualized in Figs.~\ref{FIG:3D_AXISSYMMETRIC_ONION}d and ~\ref{FIG:3D_AXISSYMMETRIC_ONION}e respectively. We note that while the observations reported in Fig.~\ref{FIG:3D_AXISSYMMETRIC_ONION} are made for a particular structure with a central bridge feature diameter of 2 nm, they hold equally for the structures we optimized for other bridge diameters.

\begin{figure*}[!]
	\centering
	{
		\includegraphics[width=0.98\textwidth]{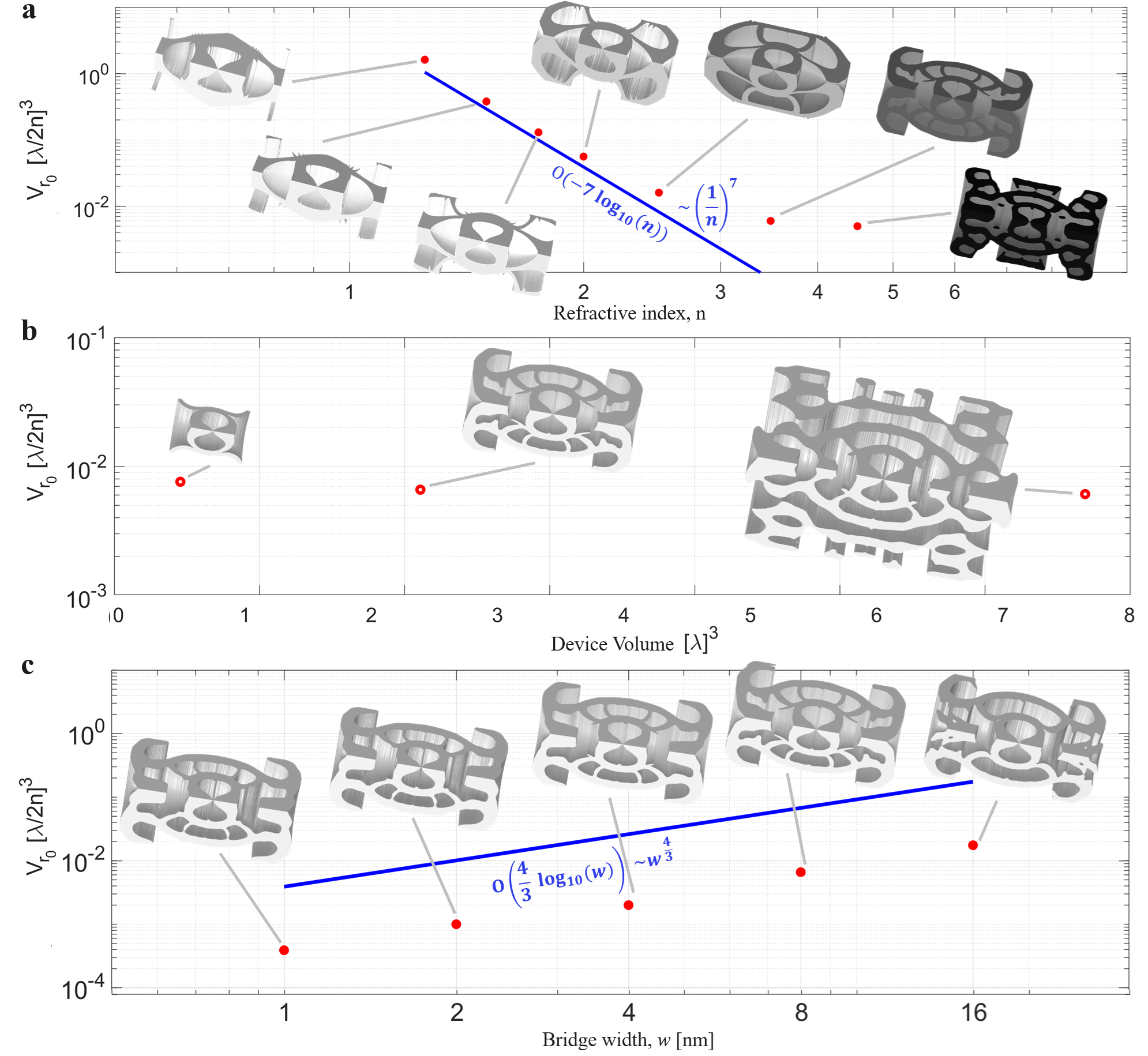} \caption{$V_{\textbf{r}_0}$ for sets of sculpted structures as a function of the \textbf{(a)} refractive index, \textbf{(b)} device volume, \textbf{(c)} central bridge width. The inserts show a 90$^o$ revolution of the optimized axisymmetric geometries. The black lines are reference lines for determining orders of proportionality. \label{FIG:PARAMETER_STUDIES_AXISYM}}
	}
\end{figure*}

\begin{figure*}[!]
	\centering
	{
		\includegraphics[width=0.95\textwidth]{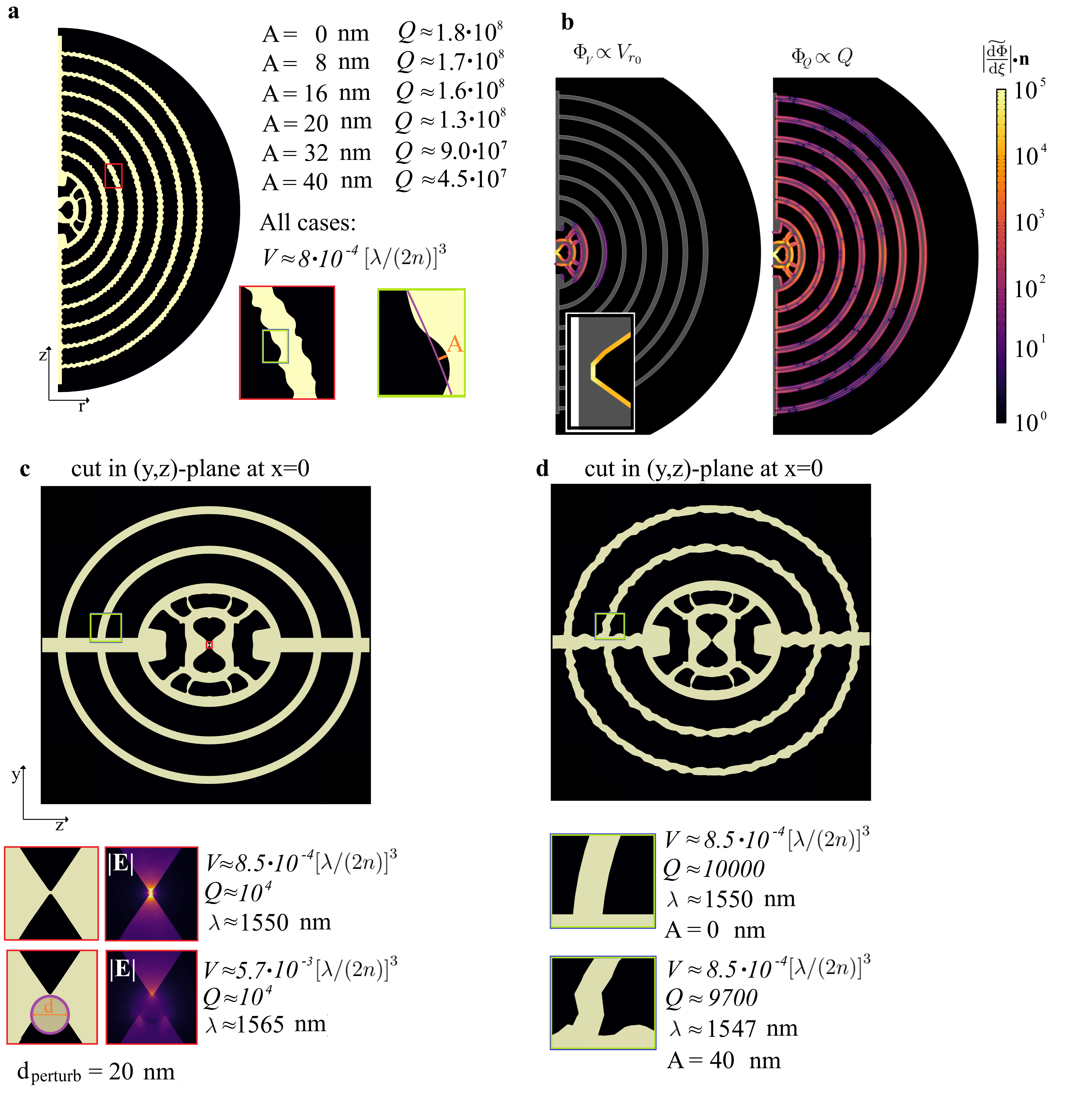} \caption{Geometric perturbation studies for the design in Fig.~\ref{FIG:PARAMETER_STUDIES_AXISYM}c. \textbf{(a-b)} Uses the axisymmetric model, while \textbf{(c-d)} uses the full 3D model. \textbf{(a)} Global perturbations of the shell boundaries, and the inserts showing a representative perturbation, with $Q$ and $V_{\textbf{r}_0}$ reported for six perturbation amplitudes. \textbf{(b)} Sensitivities towards boundary perturbations in the normal direction for $Q$ (right) and $V_{\textbf{r}_0}$ (left) with an insert showing the boundary sensitivity near the central bridge. \textbf{(c)} Local spherical perturbation to the central bridge (red box zoom), inserts show the unperturbed (top) and perturbed (bottom) center geometry (left) and field (right) along with the values for $V_{\textbf{r}_0}$, $Q$ and the resonant wavelength. \textbf{(d)} Global perturbation to the shells, with inserts (green box) showing the unperturbed (top) and a perturbed (bottom) case, along with values for $V_{\textbf{r}_0}$, $Q$, resonant wavelength and perturbation amplitude.  \label{FIG:PERTURBATIONS_STUDIES_FIG4}}
	}
\end{figure*}

\textcolor{black}{Finally, we study the sensitivity of the optimized design geometry and performance towards changes in material and geometric design limitations. First, the effect of material choice (refractive index), allowable design volume and central feature size is studied in terms of the achievable field concentration. The results of these studies are reported in Fig.~\ref{FIG:PARAMETER_STUDIES_AXISYM}. Second, the sensitivities of $V_{\textbf{r}_0}$ and $Q$ towards geometric perturbations of the particular design reported in Fig.~\ref{FIG:3D_AXISSYMMETRIC_ONION}c are studied. Here both axisymmetric and asymmetric local and global perturbation are studied and the results presented in Fig.~\ref{FIG:PERTURBATIONS_STUDIES_FIG4}.}

Firstly, seven devices are sculpted to study the effect of the choice of material refractive index. Here a central feature width of 8 nm and an overall maximum device volume of approximately $2 [\lambda]^3$ is selected, see Fig. \ref{FIG:PARAMETER_STUDIES_AXISYM}a. Remarkably, it is observed that for lower index values ($n \in [1.25,3.0]$) the achievable $V_{\textbf{r}_0}$ is strongly dependent on the refractive index, showing no less than a $1/n^7$-scaling across part of this interval. It is seen that for refractive indices corresponding to those of polymers ($n \approx 1.25$) not to be possible to achieve $V_{\textbf{r}_0} < 1 \left[\lambda/(2 n_{\text{InP}})\right]^3$, suggesting a limited utility of polymer-based 3D printing for realizing the studied devices. Further, it is seen that as $n$ reaches values above $n = 3.5$, the reduction of $V_{\textbf{r}_0}$ flattens out, revealing that once the index contrast between the background and the dielectric becomes sufficiently large ($n > 3.5$) the effect of increasing $n$ on the attainable $V_{\textbf{r}_0}$ tapers off. 

The second study concerns the effect of varying the maximally allowable device volume on $V_{\textbf{r}_0}$ (see Fig. \ref{FIG:PARAMETER_STUDIES_AXISYM}b). Here $n = 3.5$ and a central feature width of 8 nm are chosen as fixed parameters. Three device volumes spanning an order of magnitude are considered, namely $V \approx \lbrace 0.5, 2.2, 6.7\rbrace [\lambda]^3$. The overall allowable device volume is seen to only have a limited effect on the attainable $V_{\textbf{r}_0}$, only reducing by approximately 25\% as the device volume increases by more than a factor of 10. Thus, once the design volume is large enough to contain the central toroid-like feature, not much is gained in terms of the single-emitter mode-volume by increasing the design volume further. 

The third study concerns the significance of the central feature width $w$, with results presented in Fig. \ref{FIG:PARAMETER_STUDIES_AXISYM}c. Here $n = 3.5$ and a device volume of approximately $2 [\lambda]^3$ is used. From the panel, a scaling of $w^{4/3}$ is observed, revealing that until the assumed physics model breaks down, as the central feature approaches the atomic scale, the mode volume decreases with decreasing bridge width, approaching $V_{\textbf{r}_0} = 10^{-4} \left[\lambda/(2 n)\right]^3$ as $w$ approaches 0.1 nm.

\textcolor{black}{Next, considering the design in Fig.~\ref{FIG:3D_AXISSYMMETRIC_ONION}c, we study the effect on  $V_{\textbf{r}_0}$ and $Q$ of perturbing the geometry, offering a rough gauge of the designs sensitivity to fabrication imperfections.}

\textcolor{black}{First, we consider the effect of introducing axisymmetric perturbations to the six rings surrounding the central feature. The perturbations are introduced by shifting all ring boundaries by $\Delta b = A \sin(2\pi z / \Delta z) \sin(2\pi r / \Delta r)$ where the characteristic period of the perturbation $\Delta z = \Delta r$, is varied from 100 nm to 250 nm and the perturbation amplitude $A$ is varied from 0 nm to 40 nm. The electromagnetic field for the relevant quasi-normal mode is then computed using the axisymmetric model for each perturbed geometry, and $V_{\textbf{r}_0}$ and $Q$ are calculated. A cross section of a representative perturbed design is presented in Fig.~\ref{FIG:PERTURBATIONS_STUDIES_FIG4}a, along with values for $V_{\textbf{r}_0}$ and $Q$ for six representative perturbation amplitudes at $\Delta r = \Delta z = 100$ nm. Remarkably, we find that $Q$ at most drops by $30\%$ across all studied perturbations with amplitudes up to $20$ nm, a magnitude that is within the accuracy of electron-beam-lithography-based fabrication. As the amplitude grows beyond 20 nm, the reduction in $Q$ is more pronounced, approaching $90\%$ at 40 nm. We find that the mode volume (to the first two significant digits) remains independent of the perturbation amplitude and wavelength.}

\textcolor{black}{Second, again assuming the axisymmetric model, we compute the shape-sensitivity magnitude in the normal direction at all boundaries of the design in Fig.~\ref{FIG:3D_AXISSYMMETRIC_ONION}c for $V_{\textbf{r}_0}$ and $Q$. That is, we gauge the sensitive of each metric towards a perturbation of the boundary in the outward normal direction. The results are presented in Fig.~\ref{FIG:PERTURBATIONS_STUDIES_FIG4}b, showing the sensitivity of $V_{\textbf{r}_0}$ in the left panel and of $Q$ in the right, respectively. From the figure it is clear that $V_{\textbf{r}_0}$ is orders of magnitude more sensitive to perturbations near the central bridge feature, compared to elsewhere in the geometry. While $Q$ is observed to be most sensitive to perturbations close to the central feature, this metric is significantly more sensitive to perturbations at any other boundary of the design.} 


\textcolor{black}{Next, we study asymmetric perturbations, thus requiring a full 3D model. For computational feasibility we must restrict the model size and therefore study a reduced version of the design from Fig.~\ref{FIG:3D_AXISSYMMETRIC_ONION}c with only the three innermost shells included.} 

\textcolor{black}{First, we consider a spherical solid inclusion of varying diameter placed in contact with and thus perturbing the central bridge (Fig.~\ref{FIG:PERTURBATIONS_STUDIES_FIG4}c). It is found that the mode-volume increases monotonously with inclusion diameter, resulting in an increase by a factor of $\approx 6.7$ for an inclusion diameter of 20[nm] compared to the unperturbed reference. Noting that the observed mode-volume increase is slower than the scaling law reported on Fig.~\ref{FIG:PARAMETER_STUDIES_AXISYM}c, suggesting that while the asymmetric perturbation is detrimental, it has a smaller effect than increasing the bridge diameter similarly. The quality factor is found to be insensitive to this perturbation, remaining approximately $10^4$ independent of inclusion size. Meanwhile, it is worth remarking that despite $Q$ being insensitive, the frequency of the quasi-normal modes increases monotonously with inclusion diameter by a significant $\Delta\lambda\approx15$ nm for the 20 nm diameter inclusion.} 

\textcolor{black}{Second, we consider a global perturbation to the boundaries of the two outer shells by shifting these according to the function $\Delta \textbf{b} = A \sin(2\pi x / \Delta x) \sin(2\pi y / \Delta y) \sin(2\pi z / \Delta z)$ with $A \in \left[0\mathrm{nm},40\mathrm{nm}\right]$ and $\Delta x = \Delta y = \Delta z \in \left[100\mathrm{nm},250\mathrm{nm}\right]$. We find that both $V_{\textbf{r}_0}$ and $Q$ are largely insensitive to the perturbations, with the mode volume remaining constant (to the first two digits) and the quality factor dropping by at most $3\%$ for perturbation amplitudes up to 40 nm.}

\section{Conclusions}

In this work we have \textcolor{black}{shown that 3D-sculpting of dielectric cavities allow the simultaneous achievement of ultra-low optical mode volumes and ultra-high $Q$-factors}. We have found that it is possible to reduce the single-emitter mode volumes more than three (four) orders of magnitude below the diffraction limit for material (air) confinement. This constitutes a reduction of between one and two orders of magnitude relative to 2D-patterned planar counterparts proposed in recent literature. Further, we show that by introducing a set of carefully calibrated ellipsoidal shells, and a connecting rod for mechanical stability, it is possible to increase the spectral confinement by orders of magnitude ($Q > 10^8$ demonstrated numerically), thus revealing a path to systematically and individually tailor both $Q$ and $V_{\textbf{r}_0}$ depending on the desired applications. 

With expected advances in nano-scale fabrication technology, the \textcolor{black}{cavity geometries identified} in this work hold the potential for realizing light-matter interaction at unprecedented strengths in near-lossless dielectric environments. As the phenomena studied here translate near-seamlessly from telecom to microwave wavelengths, we expect experimental investigation and exploitation of the proposed geometries in the microwave regime, using existing 3D-printing tools in the near future.

\section*{Author Information}

The authors declare no competing financial interest. 

\section*{Acknowledgement}
The authors thank Olav Breinbjerg for helpful discussions and the Danish National Research Foundation for funding through grant DNRF147, NanoPhoton.

\bibliography{References}

\newpage
\clearpage
\onecolumngrid

\section{Supporting Information} \label{SEC:SUPPLE_INFO}
Details regarding the model assumed for the physics, the inverse design problem being solved and the values used for the set of free- and hyper-parameters are provided in the following.

\section*{Numerical Modeling and Inverse Design} 

All electromagnetic simulations carried out for this work assume time-harmonic field behaviour \cite{BOOK_EM_GRIFFITHS}. The physics is modeled in a spatially truncated domain $\Omega_I$ employing first-order absorbing boundary conditions on all truncation boundaries $\Gamma_{\mathrm{Rest}}$. A subset of the modeling domain is designated to be the design domain $\Omega_D \subseteq \Omega_I$, in which the structure under design resides. Specifically for the model used in the 3D-sculpting process; the full 3D modeling and design (Figs. 1-3 in the article) excite the system using a spatially truncated (Gaussian envelope), linearly polarized, planewave imposed via an additional boundary condition on part of the outer boundary denoted $\Gamma_{\mathrm{PW}}$; the axisymmetric modeling and design (Figs. 1,3-5 in the article) excite the system using a cylindrically-symmetric, spatially-truncated (Gaussian envelope) wave impinging on the device from the outer boundary. Specifically for the model used to evaluate the optimized geometries no external excitation is used, i.e. $\Gamma_{\mathrm{PW}} = Ø$. Instead the problem is solved as an eigenvalue problem in order to identify the localized eigenmode and associated eigenfrequency, which (almost perfectly) coincide with the excitation frequency used in the inverse design model (to within less than a $0.1\%$ discrepancy). Assuming Cartesian coordinates, the model equations for the electric field phasor $\textbf{E}(\textbf{r})$ may be written as, \vspace{-30pt}

\begin{eqnarray}
	\nabla \times \nabla \times \textbf{E}(\textbf{r}) - \frac{\omega^2}{c^2} \varepsilon_r(\textbf{r}) \textbf{E}(\textbf{r}) = \textbf{0}&,& \ \ \ \textbf{r} \in \boldmath{\Omega}_I \subset \mathbb{R}^3, \nonumber \\ 
	\textbf{n} \times \left(\nabla \times \textbf{E}(\textbf{r}) \right) - \mathrm{i} \frac{\omega}{c} \textbf{n} \times \left(\textbf{E}(\textbf{r}) \times \textbf{n} \right) = \textbf{0} &.& \ \ \ \textbf{r} \in \Gamma_{\mathrm{Rest}}, \label{EQN:MAXWELL_EQUATION_E_FREQUENCY_DOMAIN}  \\ \nonumber	
	\textbf{n} \times \left(\nabla \times \textbf{E}(\textbf{r}) \right) - \mathrm{i} \frac{\omega}{c} \textbf{n} \times \left(\textbf{E}(\textbf{r}) \times \textbf{n} \right) = \textbf{f}(\textbf{r}) &,& \ \ \ \textbf{r} \in \Gamma_{\mathrm{PW}}, \\ \nonumber
	\textbf{f}(\textbf{r}) = - \textbf{n} \times \left( \textbf{E}_0(\textbf{r}) \times \left(\mathrm{i} \frac{\omega}{c} €\left(\textbf{n} - \textbf{k}_{dir}\right) \right) \right) e^{-i \frac{\omega}{c} \textbf{k}_{dir} \cdot \textbf{r}}
\end{eqnarray}

\noindent Here $\omega$ denotes the angular frequency, $c$ the speed of light in vacuum, $\varepsilon_r(\textbf{r})$ the relative permittivity, $\textbf{E}_0(\textbf{r})$ denotes the magnitude and polarization of the Gaussian-enveloped incident planewave with  the propagation direction $\textbf{k}_{dir}$. The surface normal is denoted $\textbf{n}$, the gradient operator is denoted $\nabla$, the cross product is denoted $\times$ and the imaginary unit is denoted i. 

For all studies, the wavelength of the electromagnetic field is taken to be $\lambda = 1550 [\mathrm{nm}]$ ($\omega = 2\pi\frac{c}{\lambda}$). Noting here that, as the model equations themselves are scale invariant (apart from any material dispersion), the numerical value of the wavelength is not important for the conclusions drawn in the paper and thus all results in the main article are reported in terms of free-space wavelengths $\lambda$ rather than a particular numerical value. 

The model equations, eq.~(\ref{EQN:MAXWELL_EQUATION_E_FREQUENCY_DOMAIN}), are discretized on a structured grid using the finite element method \cite{BOOK_FEM_JIN} applying first-order Nedelec elements as basis functions when solving the inverse sculpting problem and second-order Nedelec elements when solving the eigenproblem to evaluated to optimized geometries. In all cases MUMPS is used as the numerical solver for the discretized model system. 

As stated in the main article the 3D-sculpting is carried out using density-based topology optimization \cite{BENDSOE_KIKUCHI_1988} to determine the material distribution in $\Omega_I$ that minimizes the single emitter mode volume for the given problem configuration. The fundamentals of the topology optimization method applied to electromagnetic problems may be understood from a tutorial paper by Christiansen and Sigmund\cite{CHRISTIANSEN_SIGMUND_COMSOL_2020} and further information may be garnered from references theirin. In brief, the material distribution in $\Omega_I$ is controlled by a design field $\xi(\textbf{r}) \in [0,1]$, which is discretized into piecewise constant voxels coinciding with the finite element mesh used to discretize the physics model. After employing a standard smoothing and thresholding procedure on the discretized design field, it is used to linearly interpolate the relative permittivity $\varepsilon_r(\textbf{r})$ between a background material for $\xi(\textbf{r}) = 0$ and the device material for $\xi(\textbf{r}) = 1$ in each voxel. The sculpting procedure is iterative in nature, where $\xi(\textbf{r})$ is changed to minimize $\Phi$ using the gradient-based globally convergent Method of Moving Assymptotes \cite{SVANBERG_2002}, employing a maximum of 3 inner iterations per outer design iteration. The gradient information is obtained using adjoint sensitivity analysis \cite{TORTORELLI_ET_AL_1994}. In order to promote a well behaved optimization process and a final physically admissible design, and to allow for the imposition of length-scale in the design, a pamping \cite{JENSEN_SIGMUND_2011}, smoothing, thresholding \cite{WANG_ET_AL_2011} and continuation procedure is employed. 

Each 3D-freeform sculpting study is executed with a total of 500 design iterations, distributed over 10 continuation steps, using the parameter values listed in the first line of Tab. \ref{TAB:OPTIMIZATION_PARAMETERS_3D}. The axisymmetric design studies are executed employing a total of 200 design iterations using four continuation steps with the parameter values listed in second line of Tab. \ref{TAB:OPTIMIZATION_PARAMETERS_3D}.

We note that all cases are started using a single uniform initial guess for the design field, namely $\xi = 0.5 \ \forall \ \textbf{r} \in \Omega_D, \ \xi = 0 \ \forall \ \textbf{r} \in \Omega_I \backslash \Omega_D$. It is stressed that no issues of the algorithm converging to poorly performing local minima are observed in this study regardless of the design problem considered. That is, only a single initial guess was needed for each design and thus no multi-start procedure or the like is employed.
 
\begin{table}[h!]
	\centering{
		\begin{tabular}{cccccc}
			$\eta$ & $\beta$ & $\alpha_P$ & $r_f$ & $n_{\mathrm{iter}/\beta}$ \\ \hline
			\\ [-1em]
			0.5 & $\lbrace 1, 2, 4, 8, 16, 32, 64, 64,  64,  64 \rbrace$ & $\lbrace 0.05, 0.05, 0.1, 0.1, 0, 0, 0, 0, 0, 0 \rbrace$ & 50 [nm] & 50 \\
			0.5 & $\lbrace 5, 10, 20, 40 \rbrace$ & $\lbrace 0.05, 0.025, 0.01, 0.0 \rbrace$ & 30 [nm] & 50 \\
			
		\end{tabular}
		\caption{Parameters used for the pamping, smoothing, thresholding and continuation. The threshold level $\eta$, threshold strength $\beta$, pamping magnitude $\alpha_P$, filter radius $\textbf{r}_f$ and fixed number of design iterations $n_{\mathrm{iter}/\beta}$ for each continuation step.} \label{TAB:OPTIMIZATION_PARAMETERS_3D}
	}
\end{table}

The models used in both the inverse design and evaluation processes were all implemented utilizing COMSOL Multiphysics \cite{COMSOL61main}. In the discretized numerical model the design field $\xi(\textbf{r})$ is itself discretized using elementwise constant design variables on the structured FEM mesh used for solving the physics problem. For the 3D-freeform design study, brick elements with a maximum side-length of 25 nm are used throughout $\Omega_I$, except near $\textbf{r}_0$ where a fixed mesh of cubic elements with 10 nm side-length is used to ensure modeling accuracy. The design discretization is purposefully kept identical across the 2D-patterning and 3D-sculpting cases to faciliate directly comparible results. This relatively coarse design descretization is dictated by the computational complexity associated with the numerical solution of the model equations in eq.~(\ref{EQN:MAXWELL_EQUATION_E_FREQUENCY_DOMAIN}) for the largest device dimensions considered in the study, requiring approximately $0.5\cdot 10^6$ elements to discretize the model problem. For the axisymmetric studies a design resolution of 0.5 nm is used in the region within 20 nm of $\textbf{r}_0$, while a maximum resolution of 10 nm is used in the region further from $\textbf{r}_0$ using a gradually coarsening non-uniform mesh.

For the fully 3D-freeform design studies the model domain is taken to be cubic shaped $\Omega_I = \Delta x_I \otimes \Delta y_I \otimes \Delta z_I = [-2 \lambda, 2 \lambda]^3$. The design domain $\Omega_D$ is taken to be brick-shaped with in-plane side lengths of two free-space wavelengths. This choice is based on an investigation (not shown) revealing that the attainable $V_{\textbf{r}_0}$ is nearly unaffected by geometric variations further than $\lambda$ from $\textbf{r}_0$. Four different configurations of $\Omega_D$ are studied. Namely; a reference configuration with a fixed device geometry in the out-of-plane direction taking $\Omega_D = \Delta x_D \otimes \Delta y_D \otimes \Delta z_D = [-\lambda, \lambda]^2 \otimes [-100 \mathrm{nm}, 100 \mathrm{nm}]$; and three configurations with design freedom in all spatial dimensions for increasing device thicknesses, $\Delta z_D \in \lbrace[-100 \mathrm{nm}, 100 \mathrm{nm}], [-400 \mathrm{nm}, 400 \mathrm{nm}], [-800 \mathrm{nm}, 800 \mathrm{nm}]\rbrace$. Finally, Mirror-symmetry is imposed normal to the (x,z)- and (y,z)-planes about $\textbf{r}_0 = (0,0,0)$ to reduce the computational cost associated with solving the model problem. The material parameter of the design and background are chosen to correspond to Indium Phosphide ($n_{InP} = 3.17$) and air ($n_{BG} = 1.0$) respectively, with a cubic region with 10 nm sidelength around $\textbf{r}_0$ fixed as air, thus investigating the achievable air-confinement of the electromagnetic field.

For the axisymmetric design studies, the model domain dimensions are taken to be $\Omega_I = \Delta r_I \times \Delta z_I = [-1.5 \lambda, 1.5 \lambda]^2$, with varying rectangular $\Omega_D$ sizes, device volumes, refractive indices and minimum feature sizes. Here both air and solid confinement is studied by fixing a small cylindrical region around $\textbf{r}_0$ to be either air or solid, with the radius of this region defined by the minimum allowable feature size as stated for each study. The study of both cases is performed, as it has previously been found that the achievable single-emitter mode volume when confining the field inside the solid is significantly larger than when confining the field in air.

\end{document}